\documentclass[10pt,a4paper,twocolumn]{article}
\usepackage[latin1]{inputenc}

\usepackage[caption=false,font=footnotesize]{subfig}

\usepackage{amsmath}
\usepackage{amsfonts}
\usepackage{amssymb}
\usepackage{graphicx}
\usepackage{color,enumerate,cite,cases}

\newcommand{\R}{R_{xy}}
\newcommand{\Rxx}{R_{xx}}
\newcommand{\C}{\hat{C}_{xy}}
\newcommand{\Cxx}{\hat{C}_{xx}}

\newcommand{\F}[1]{\mathcal{F}\left(#1\right)}
\newcommand{\Fi}[1]{\mathcal{F}^{-1}\left(#1\right)}
\newcommand{\expec}[1]{ \mathcal{E}\left\{ #1 \right\}  }	
\newcommand{\ii}{ \mathrm{i} }

\usepackage{authblk}

\date{}                    
\begin{document}
	\title{Unbiased two-windows approach for Welch's method} 	
	\author{Eduardo Martini
		\\
        D\'epartement Fluides, Thermique et Combustion, Institut Pprime, CNRS, Universit\' e de Poitiers, ENSMA, 86000 Poitiers, France  	
        \\
        eduardo.martini@univ-poitiers.fr}
	
	\maketitle
	\begin{abstract}
		Periodogram methods are widely used for the estimation of power- and cross-spectra, of which  Welch's method is the most popular. Previous studies have analyzed the variance of the power spectra estimates and developed analytical probability functions, showing that the approach is unbiased when applied to white-noise signals or in the limit of infinite window lengths. However,  no explicit expression for the estimation bias is available for more complex signals when finite windows are used.  In this study, we show that, for finite window lengths,  Welch's method is biased for all signals other than the white-noise signal. A novel two-window approach that is unbiased when applied to signals with bounded correlation lengths is proposed. Numerical experiments are used to illustrate the advantages of the novel approach. 
		\end{abstract}
	\section{Introduction}
	
	Power spectra estimation (PSE) has received considerable attention in the last decades. Two main classes of estimators are typically used, parametric and non-parametric. 
	
	Parametric estimators estimate parameters in a pre-defined model to represent the signal power spectra, while non-parametric models do not make assumptions about the data. The first non-parametric model proposed was the periodogram approach, proposed by \cite{schuster1898investigation}, which was improved in several studies \cite{proakis2001digital,bartlett1948smoothing,stoica1997introduction}. The most well-known derivation of the periodogram approach is Welch's method \cite{welch1967use}. 
	
	In its first use, the periodogram was based on Fourier transforming all the available data, using the magnitude of coefficients as a PSE. The approach is asymptomatically unbiased in the limit of large sample lengths, but the estimate has a large variance. To reduce the variance, a signal can be divided into blocks, and the PSE estimation on each block is averaged. For a given signal length, there is a trade-off between the number of blocks and the block length, which corresponds to a trade-off between the variance and the spectral resolution of the PSE.   Welch's method alleviates this trade-off by overlapping blocks, allowing larger blocks and/or more samples to be used.
	The use of the fast-Fourier transform (FFT) algorithm makes these approaches computationally inexpensive, and they are thus widely used to estimate not only frequency-domain statistics, e.g.,  power and cross spectra, but also time-domain statistics, e.g., auto- and cross-correlations, which are obtained from an inverse Fourier transform of their frequency-domain counterparts. 
	
	Despite its popularity, few studies have extended the analysis of Welch. The variance trends for increasing overlapping were studied in \cite{bronez1992performance,jokinen2000windowing}, and the use of a circular PSE was proposed by \cite{barbe2010welch}, leading to an equal weighting of all the available data. Probability-density functions for the estimates using  Welch's method were obtained by \cite{johnson1999probability}. Analytical results have shown that the estimates are unbiased for noise-white signals, and numerical results showed small biases for a few other signals considered.
	
	In this work, biases in spectral estimations are studied converting them back to the time-domain. An explicit expression for the estimation bias is derived, showing that the estimation of any signal other than the white-noise signal is biased when finite window lengths are used. A method using two windows is proposed and shown to eliminate estimation bias if the cross-correlation function is bounded.
	
    This work is structured as follows.  The definitions of the cross-correlation/cross-spectral density and Welch's method are reviewed in section \ref{sec:Welch}. A novel, two-window approach, is proposed in \ref{sec:twowindows}. Section \ref{sec:num_tests} presents numerical tests, and conclusions are draw in section \ref{sec:conclusions}.

\section{Revisiting  Welch's method} \label{sec:Welch}
			
	The cross-correlations, $ \R $, between two stochastic stationary signals, $ x $ and $ y $, assumed to be complex-valued, is defined as 
	\begin{align}\label{eq:R}
		\R(\tau) =  \expec{ x(t) y^*(t-\tau) },
	\end{align}
	where $ \expec{ \cdot } $ represents the expected value and $ ^* $ the complex conjugation. 
	The cross-spectral density, $ \C $,  is defined as 
	\begin{align}\label{eq:C}
	\C(\omega) = \F{\R(\tau)},
	\end{align}
	where 
	\begin{equation}\label{eq:Fourier}
		\hat f(\omega)=\F{f(t) }  = \int_{-\infty}^{\infty}f(t) e^{-\ii \omega \tau} d\tau,
	\end{equation}
	is the Fourier transform. That is, $ \R$ and $\C $ are different expressions of the same information in the time and frequency domains, respectively. 


	For stationary signals, the averaging in \eqref{eq:R} is typically taken as a time average. Computing $ \R $ thus requires a convolution between two signals, which can be an expensive operation. To avoid these costs, Welch's method \cite{welch1967use} is typically used to estimate $\C$ directly. The method divides the signal into blocks, estimates $\C$ on each of these blocks using the periodogram approach, and then averages these estimates. The use of fast-Fourier transforms (FFT) greatly reduces the computational costs when compared to a direct convolution on the time domain. 


	Harris~\cite{harris1978use} studied the effect of using different windowing functions on the spectral estimates. Using a windowing function $ w (t)$,  $ \C $ is estimated as 
	\begin{align}\label{eq:Cp}
		\C'(\omega) 
		&=\expec{ \hat{x}_w(\omega) \hat{y}_w^*(\omega) } 
	\end{align}
	where
	\begin{equation}\label{eq:hatwelch}
		\hat{x}_w(\omega) = \int_{-\infty}^\infty x(t) w(t) e^{\ii \omega t} dt,
	\end{equation}
	is the Fourier transform of the windowed signal, with an analogous expression for $ \hat{y}_\omega $. 
	
	Note that since $ \hat{x} $  is not not well defined, as $ x $ is a stochastic and stationary signal, it is not square integrable, and thus $ \expec{\hat{x}\hat{y}} $ an ill-defined expression.  This motivates the definition of $\C $ as in \eqref{eq:C}. As $ w(t) $ is zero outside each block, $ \hat{x}_w $ and $ \hat{y}_w $, and thus also \eqref{eq:Cp}, are well defined. Harris investigated the impact of different windows on the estimate's spectral resolution and spectral leackage.

    \subsection{Bias}
	To explore biases in the method, \eqref{eq:C} is inverted to obtain the time-domain representation of the frequency-domain statistics, i.e., an inverse Fourier transform is applied to $ \C' $, 
	\begin{align}\label{eq:Rp}
		\begin{aligned}
		\R'(\tau)     &= \Fi{ \C'(\omega)}.
		\end{aligned}
	\end{align}
	Using \eqref{eq:Cp}, \eqref{eq:Rp} is re-written as
		\begin{align}
		\begin{aligned}
				\R'(\tau)  &=  \int_{-\infty}^{\infty} \expec{ x(t')y^H(t'-\tau)} w(t')w(t'-\tau)  d t'  \\
								&=  \R(\tau )\int_{-\infty}^{\infty}w(t')w(t'-\tau) d t' ,
		\end{aligned}
		\end{align}
		and thus		
		\begin{align}
		\label{eq:bias}
		\begin{aligned}
			\R'(\tau)     
			&=  \R(\tau) W(\tau), 
		\end{aligned}
	\end{align}
	where 
	\begin{align} \label{eq:W}
		W(\tau) = \int_{-\infty}^\infty  w(t) w(t-\tau)) dt.
	\end{align}

\begin{figure}[t!]
		\centering
		\includegraphics[scale=0.5]{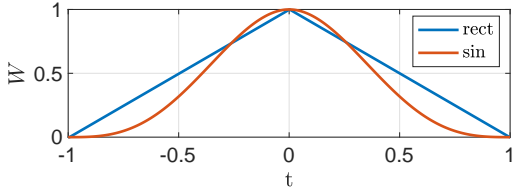}
		\caption{Super windows for two common windows used with Welch's method.}
		\label{fig:wwelch}
	\end{figure}

	Equation \eqref{eq:bias} shows that Welch's method reproduces the cross-correlation function modulated by $ W(\tau) $, which will be refered to as a super-window. The estimation is unbiased only if $\R(\tau)W(\tau)=\R(\tau)$, i.e., if $ W(\tau)=1 $ for all $ \tau $ where $ \R(\tau)\neq 0 $. As $ W(\tau) $ is obtained as a convolution of the windowing function with itself, it is equal to one only at $\tau=0$, as illustrated in  figure~\ref{fig:wwelch}. As a consequence, Welch's method is unbiased when applied to a white-noise signal, as shown by \cite{johnson1999probability}. However, for any other signal, Welch's method is biased, under-predicting $ \R $.

	From Parseval's theorem, the $ L^2 $ norm of the estimation bias in the time and frequency domains is the same, i.e., 
	\begin{equation}\label{eq:error}
		\begin{aligned}
			e^2 = & \int |\R(\tau)-\R'(\tau)|^2 d\tau \\
			= &\int |\C(\omega)-\C'(\omega)|^2 d\omega.
		\end{aligned}
	\end{equation}
	An analytical expression for the frequency-domain estimate reads
	\begin{equation}\label{eq:biasfreq}
		\begin{aligned}
			\C'(\omega) &= \C(\omega) * \hat{W}(\omega)
		\end{aligned}
	\end{equation}
	where $ \hat{W}(\omega) =| \hat{w}(\omega)|^2 $, as recognized by Welch~\cite{welch1967use}. However, from \eqref{eq:biasfreq}, it is not clear what are the biases, e.g., if the CSD is under- or over-estimated, and does not suggest a strategy for reducing them.  

	The present analysis suggests that estimation biases can be limited if the super-window is flat in the region where $\R(\tau)\neq0$. 
    As figure \ref{fig:wwelch} suggests, this is not possible using a single window. Note however that $ W(\tau)\to 1 $ everywhere as the window length is increased, which is again consistent with the fact that periodogram approaches are unbiased in the limit of infinite block lengths. However, this convergence is typically slow and requires large windows to be used, reducing the number of blocks that can be obtained from a given data series and thus increasing the estimation variance. In the next section, a two windows approach that provides unbiased estimations with finite window lengths is proposed.

	\section{The two-windows approach}\label{sec:twowindows}
	To generalize the method to obtain an unbiased estimation for a large class of signals, different windowing functions for each signal are used. The cross-spectra is now estimated as 
	\begin{equation}\label{eq:Cp_twowin}
			\C'(\omega) 
			 =\expec{ \hat{x}_{w_l}(\omega) \hat{y}_{w_r}^*(\omega) } ,
	\end{equation}  
    where $ w_l $ and $ w_r $ are windowing functions used for $ x $ and $ y $, respectively.
	The expected value of the time-domain estimate is again given by \eqref{eq:Rp}, with 
	\begin{align} \label{eq:W_twowind}
		W(\tau) = \int_{-\infty}^\infty  w_l(t) w_r(t-\tau)) dt.
	\end{align}
	
    Again, the goal is to construct $W(\tau)$ such that $\R(\tau)=W(\tau)\R(\tau)$. For such we define $ \Omega $ as the region where $ \R(\tau)\ne 0  $, and, throughout this work, assume it to be compact.  Although this assumption excludes some types of signals, e.g., periodic signals, it contemplates most of the physically relevant scenarios. 
	By properly choosing $ w_l $ and $ w_r $, a super-window that has a value of one over $ \Omega $ can be constructed. For simplicity we assume that  $ \Omega=[\tau_c-\Delta\tau/2,\tau_c+\Delta\tau/2] $, i.e., an interval of size $ \Delta\tau $ centered at $ \tau_c $. Generalization to more complex domains is presented in section \ref{sec:ultipleOmegas}.
	
	Two different window pairs will be explored here:
	\begin{enumerate}[(a)]
		\item Rectangular windows 
		\item Infinitely smoothed windows 
	\end{enumerate}

	In (a),  $ w_l(t) $ and $ w_r(t) $ are constructed from a rectangular window,
	\begin{align}
		\label{eq:wrect}
		w_{rect}(t) & =
		\begin{cases} 
			1 , & \tau \le 1/2  \\
			0 , & \tau > 1/2  
		\end{cases},
	\end{align} 
	as
	\begin{align}
		\label{eq:wl}
		w_l(t) &= w_{rect}\left( \dfrac{t}{L}  \right), \\
		\label{eq:wr}
		w_r(t) &=\dfrac{1}{L} w_{rect}\left( \frac{t-\tau_{c}}{L+ \Delta \tau } \right),
	\end{align}
	where $ L $ is the effective window length. The result super-window is given by
	\begin{equation}\label{key}
	W(\tau) = \begin{cases}
		1 & , |\tau'| \le \Delta\tau/2 \\
		1-\dfrac{|\tau'-\frac{\Delta \tau}{2}|}{2L+\Delta \tau}
		& , \dfrac{\Delta\tau}{2} < |\tau'| \le 2L +\Delta \tau\\
		0& ,\text{ otherwise}
	\end{cases},
\end{equation}
	where $ \tau'=\tau-\tau_c$.

    The resulting super-window is flat over  $ \Omega $, with a linear decay to zero over a distance of $ 2L+\Delta \tau $. It is a $ C^0 $ function. PSE using this window pair is unbiased but will invariably contain statistical noise modulated by the super-window, thus exhibiting significant spectral leakage due to the slow decay of the super-window spectra.
	
	To minimize the spectral leakage, an infinitely-smooth super-window is sought in (b). For the signal $ x $, a rectangular $w_l$ is used, as in \eqref{eq:wl}, while for $ y $
	\begin{align}		\label{ew:wrcinf}
			w_r(t) &= \dfrac{1}{L}w_{C_\infty }\left( \frac{t-\tau_{c}}{L+ \Delta \tau } \right), 
	\end{align}
	is used, where
	\begin{subnumcases}{\label{eq:wcinf}
			w_{C_\infty}(t)=}
		1 &,  for $t <  \frac{1}{2}$ \\
	 f\left(\dfrac{t-\frac{1}{2}}{\delta_1}\right)    &, for $0<|t|-\frac{1}{2}<\delta_1$ \\
   0  &, otherwise
	\end{subnumcases}
		and
	\begin{align}
		f(t) =&	 \frac {1 +  \tanh\left(\frac{\cot\left( \pi (|t|-\frac{1}{2})  \right) }   {\delta_2}   \right) }{2},
	\end{align}
	is a function that provides an infinitely-smooth transition between 0 and 1.

    The parameter $ \delta_1 $ controls the transition's length, and $ \delta_2 $ its sharpness. In the limit of ${ \delta_2\to \infty} $, a rectangular window with length $ 1+\delta_1 $ is obtained, and when $ \delta_2\to 0 $ a linear transition is recovered. 

\begin{figure}[t]
        \centering
        \includegraphics[scale=0.5]{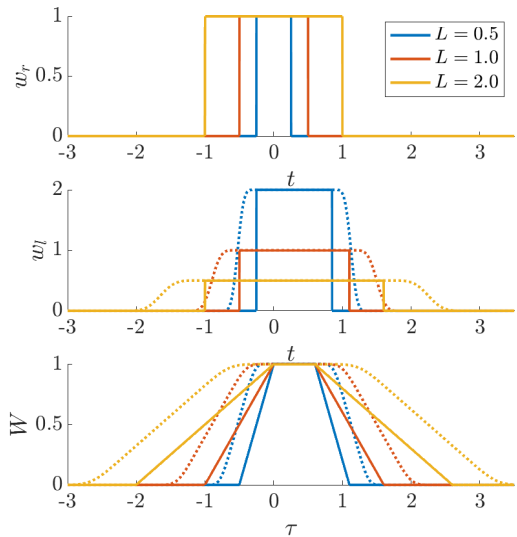}
        \caption{Illustration of $ w_l,w_r$ and $ W $ for the $ C_1 $ (solid lines)  and $ C_\infty $ (dashed lines) two-windows approach.  Here  $ \delta_1=\delta_2=0.5 $ were used. 	}
        \label{fig:windowsl} 
    \end{figure}

	Figure \ref{fig:windowsl} illustrates the different window pairs and their corresponding super-windows for different values of the parameter $ L $.  For both pairs of windows, the super-window is exactly one in $ \Omega $. The window pair (i) goes faster to zero, while (ii) has a longer, but smoother, transition. The total length of the super-windows are $ 2L+\Delta \tau $ and $ 2L + \Delta \tau + 2\delta_1(L+\Delta\tau) $ for (a) and (b), respectively.
	
\begin{figure}
		\centering
		\includegraphics[scale=0.5]{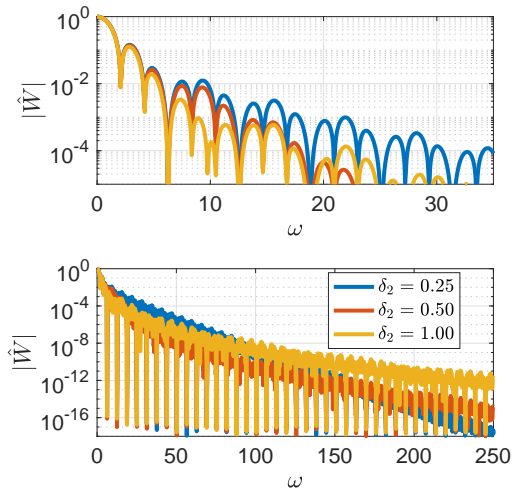}
		\caption{Super-window spectra for the $ C_\infty $ super-windows, for $ L=1$ and  $\delta_1=0.5$.}
		\label{fig:superwindows}
	\end{figure}

    Figure \ref{fig:superwindows} shows the spectra of the resulting super-window using $ L=\delta_1=1 $. The parameter $ \delta_2 $ shows a trade-off between the asymptotic decay rate of window spectra and the first side-lobe peaks.  As discussed in a previous work~\cite{emartiniIEEE_ArXiV}, infinitely smooth functions exhibit exponentially decaying spectra and are thus useful to minimize spectral leakage and, for sampled data, provide faster convergences of Fourier transforms with the sampling frequency.

\begin{figure}[t]
		\centering
		\includegraphics[scale=0.5]{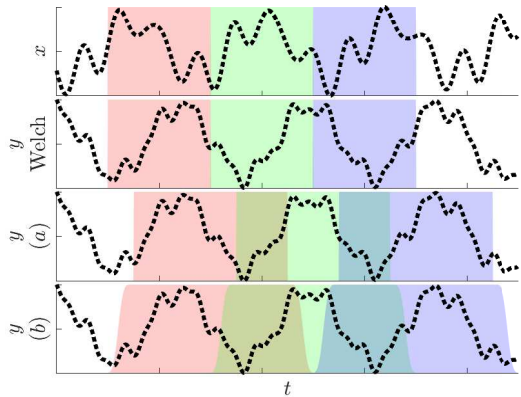}
		\caption{Illustration of the data blocks used for the $ x $ and $ y $ signals with Welch's method and the proposed approach. Fourier transforms of blocks with the same colours are  multiplied and averaged out in each method.}
		\label{fig:methodsillustration}
	\end{figure}

    An illustration of the blocks used by Welch's method and the current approach using (a) and (b) are illustrated in figure~\ref{fig:methodsillustration}. Note that although there is an overlap of the $ w_r $ windows, the PSE of each block pair are independent: as there is no overlap in $ w_l $, no pair of $ x(t) y(t+\tau)$ is used more than once, and thus there is no repeated information on different blocks.
 	
    The effective window length provides a trade-off between the computational cost, memory requirement, and super-window transitions:  larger values of $ L $ make use of FFTs to efficiently performs computation, at the price of higher memory requirement and larger super-window transitions. As the approach is unbiased for any $ L $, the dominant factor in the choice is the trade-off between CPU and memory costs.

\subsection{Estimating the region with non-zero cross-correlation, $ \Omega $} \label{sec:estOmega}

	The construction of the window pair assumes that the region $ \Omega $ is known. However, this information is often not available. In this section, we propose an approach to estimate it.
	
	As both $ x $ and $ y $ are stochastic signals, so is $ x(t)y(t+\tau) $. The expected value of the latter provides $ \R(\tau) $, as defined in \eqref{eq:R}. The domain $ \Omega $ can thus be estimated by testing if $ \R(\tau) $ is null or not. The probability density function of a product of variables has been the subject of study of several works \cite{craig1936frequency,nadarajah2016distribution,cui2016exact}, but as $ \R(\tau) $ is obtained as the average of many such products, the central limit theorem guarantees that it is asymptotically normally distributed. Defining $ \sigma $ as the standard variation of $  x(t)y(t+\tau)  $,  
	\begin{equation}
		R'_N(\tau)\to N(R(\tau),\sigma_N(\tau)) ,
	\end{equation}  
	for large $ N $, where $ R_N(\tau) $ is the estimation of $ R_N(\tau) $ from $ N $ realizations, and $ \sigma_N(\tau) =\sigma(\tau)/\sqrt{N} $ is the standard deviation of this estimate if the samples are uncorrelated.  
	
    Here the Student's T-test is used test the hypothesis that $ \R(\tau)=0 $, i.e. $ \tau\in\Omega $, with a confidence level $p$.
    
    First, an estimate of the mean and the standard deviation of $\R(\tau)$ are computed from its estimation on the i-th, $\R^{(i)}(\tau)$, as
	\begin{align}\label{key}
		\mu(\tau) &= \sum_{i=1}^{N} \R^{(i)}(\tau) ,\\	
		s(\tau) &= \sum_{i=1}^{N} \dfrac{(\R^{(i)}(\tau)-\mu(\tau))^2	}{N-1},
	\end{align}
    used to compute the t-value
	\begin{equation}\label{key}
		t(\tau) = \frac{ \mu(\tau) }{s(\tau)}.
	\end{equation}
	 
	The hypothesis that $ \R(\tau)=0 $ can be rejected, with confidence level $ p $, if $ |t|>t_c(p,N) $, where $ t_c $ is obtained from Student's T distribution~\cite{student1908probable}, for a given confidence level $ p $ and  the $N$ samples.
	
	Alternatively, a tolerance can be prescribed, i.e., $ \Omega $ is estimated as the region where we can reject the hypothesis that $ |\R(\tau)|\le\epsilon $.  The domain $\Omega$ is estimated as  the region for which the hypotheses $ \R(\tau)-\mu_0=0 $ cannot be discarded for any $ \mu_0 \in [-\epsilon,\epsilon]$. The t-value is now computed as	
	\begin{equation}\label{key}
		t(\tau,\mu_0) = \frac{ \mu-\mu_0 }{s},
	\end{equation}
	and the hypothesis that $ |\R(\tau)|\le\epsilon $ is discarded if 
	\begin{equation}\label{key}
		  \min_{\mu_0\in[-\epsilon,\epsilon]}(t(\tau,\mu_0))>t_c(p)
	\end{equation}
	
	This can be computed effortlessly by nothing that 
	\begin{subnumcases}{
		  \min_{\mu_0\in[-\epsilon,\epsilon]}(t,\tau(\mu_0)) =  }
			0 &, for $|\mu|\le\epsilon $\\
	\left|\frac{ \mu }{s} \right| - \left|\frac{  \mu_0 }{s} \right|  &, for $|\mu|>\epsilon$ 
	\end{subnumcases}
	
	In practice, this test can be applied using a downsampled time series, and thus adds little to the total computational costs.
    In this study, we use a threshold value  $ p_c $ that corresponds to a probability of  $ 99\% $ confidence using the two-sided Student's T-test.

\subsection{Enforcing symmetry when $ x=y $} \label{sec:est_autocorr}
    For the particular case where $ x=y $,  $ \Rxx=\Rxx^H $ and $ \Cxx=\Cxx^H $, where $^H$ represents the complex-conjugate transpose.  Although for properly designed window pairs the expected value of the estimation reproduces $ \R $ and $ \C $, thus reproducing their symmetries upon complex conjugation, this is only achieved asymptotically for a large number of samples. Note that  \eqref{eq:Cp}, when used for identical signals,  
    \begin{align}\label{eq:Cp_gen}
        \Cxx'(\omega) 
        &=\expec{ \hat{x}_w(\omega) \hat{x}_w^H(\omega) } ,
    \end{align}
    enforces the symmetry of $ \Cxx $ for any, however unconverged estimation, but this is not the case for \eqref{eq:Cp_twowin}, 
    \begin{equation}\label{eq:Cp_twowin_gen}
        \Cxx'(\omega) 
        =\expec{ \hat{x}_{w_l}(\omega) \hat{x}_{w_r}^H(\omega) } .
    \end{equation}  
    
    To obtain an estimator that respects this symmetry, \eqref{eq:Cp_twowin_gen} can be modified to
    \begin{align} \label{eq:Csnap}
        \Cxx'(\omega) = \expec{ \dfrac{\hat{x}_{w_l}\hat{x}_{w_r} ^H + \hat{x}_{w_r}\hat{x}_{w_l} ^H }{2}},
    \end{align}
    thus ensuring invariance w.r.t complex-transposition.

\subsection{Approach for $ \Omega $ made of several intervals} \label{sec:ultipleOmegas}
    
    Previously, it was assumed that $ \Omega $ consists of a single interval. Here we discuss approaches when it is composed of several non-connected intervals. For simplicity, we assume that $ \Omega = \Omega_1 \cup \Omega_2 $, i.e., it is the union of two simply connected, disjoint, sets. The generalization to more sets is straightforward.
		
    Two distinct scenarios are possible. If $ \Omega_1$ and $ \Omega_2$ are sufficiently distant, they can be identified separately, each with a custom pair of windows. After adding both estimates, it is straight-forward to show that \eqref{eq:bias} becomes
    \begin{equation}\label{key}
        \R'(\tau)   =  \R(\tau) (W_1(\tau)+W_2(\tau)),
    \end{equation}
    where $ W_{1,2} $ are the super-windows corresponding to each window pair. The total super-window for the approach is $ W=W_1+W_2 $.

\begin{figure}[t!]
        \centering
        \subfloat[]{\includegraphics[scale=0.5]{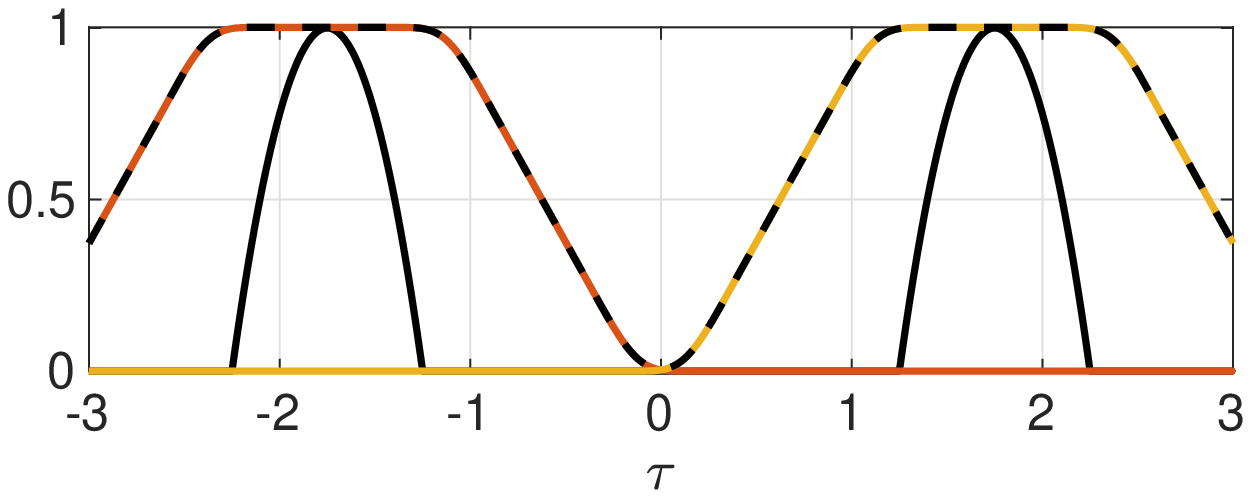}}

        \subfloat[]{\includegraphics[scale=0.5]{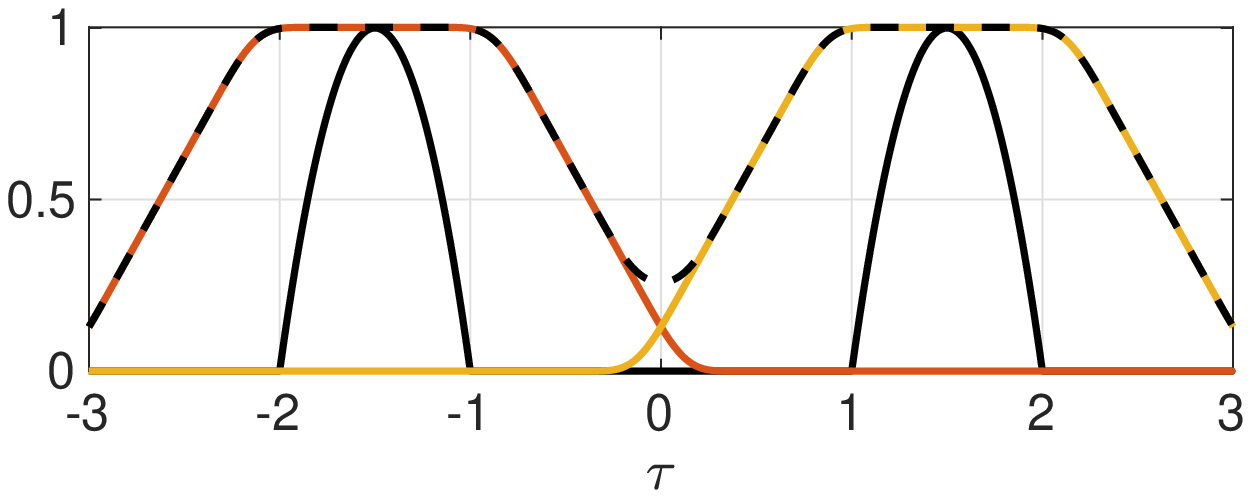}}

        \subfloat[]{\includegraphics[scale=0.5]{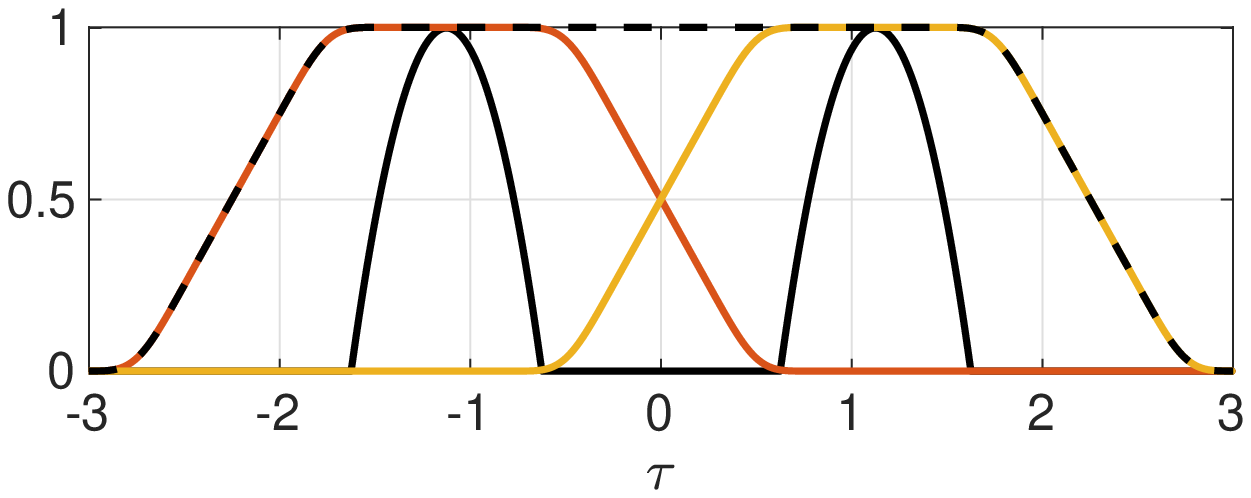}}

        \subfloat[]{\includegraphics[scale=0.5]{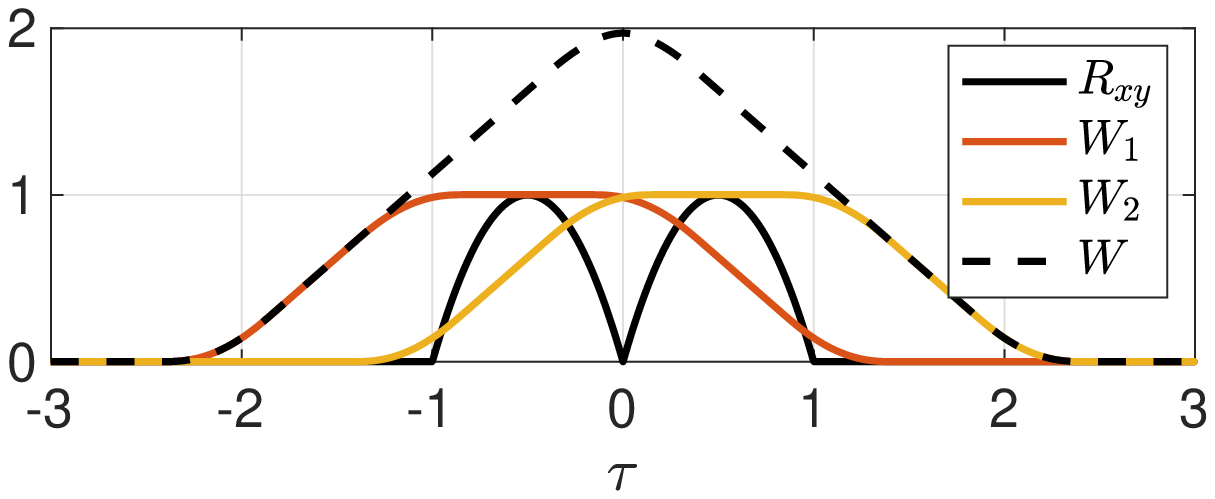}}
        \caption{Illustration of super-windows resulting from the estimation using two pairs of windows. Overlaps as in (d) should be avoided.}
        \label{fig:twowindowpairs}
    \end{figure}

    Super-windows resulting from the use of two different window pairs for progressively closer $ \Omega_1 $ and $ \Omega_2 $ are illustrated in figure \ref{fig:twowindowpairs}. If the sets are close, two approaches can be used: the different pairs can be constructed to form a larger plateau, as in figure \ref{fig:twowindowpairs}c, which is possible due to the symmetry of its tails, or a single window can be used for both sets. The scenario in figure \ref{fig:twowindowpairs}d should be avoided: overlap between super-windows plateaus and transitions introduces biases in the estimation of $ \R $, which is overestimated if $ W(\tau)>1 $. 

\section{Numerical tests}\label{sec:num_tests}
	 
    Here the proposed two-windows method is investigated numerically. For such, synthetic $ x $ and $ y $ signals are constructed from a normally distributed, white-noise, signal $s$ and two functions $r_1$ and $r_2$ as
 	\begin{align}
		x &= s*r_x, \\
		y &= s*r_y.
	\end{align}
	with $ r_x $ and $ r_y $ given by
	\begin{align}
	 	r_x(t) &= \begin{cases}
	 		1-|t/\delta| &,t<\delta \\
	 		0 &,t>\delta
	 	\end{cases} 
	 	,\\ 
	 	r_y(t) &= \begin{cases}
	 		1-|(t-\Delta t)/\delta| & , t<\delta \\
	 		0 & , t >\delta
	 	\end{cases}. 
	\end{align}
	where $ \delta=0.15 $ and $ \Delta t=0.30 $ were used.
	The resulting signals have cross-correlation given by $ \R(\tau)=r_x(\tau)*r_y(\tau) $, with $ \R(\tau)\neq0 $ for $\tau \in [-0.3,0.9] $.
		
	For illustration purposes, we investigate four different approaches to estimate $ \R $ and $ \C $:
	\begin{enumerate}[(i)]
		\item Welch's method, using a rectangular window,
		\item using rectangular windows of the same size, with a temporal shift of $ \Delta t $,  
		\item the two windows approach using rectangular windows,
		\item the two windows approach using an infinitely smooth window.
	\end{enumerate}  

\begin{figure}[t!]
		\centering
		\includegraphics[scale=0.5]{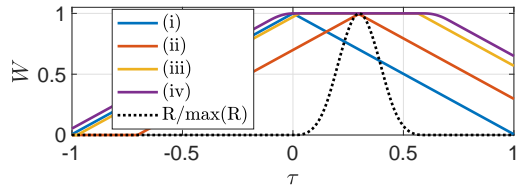}
		\caption{illustration of super-windows obtained different strategies and the normalized cross-correlation $R$.}
		\label{fig:W} 
	\end{figure}

	Approach (ii) is an intermediary between Welch's method and the two-window approach and is included to illustrate the role of translating the second window in time, which centers the super-window at $ \Omega $; and using different window sizes to create a plateau in $ \Omega $. For (i) and (ii), rectangular windows of length $1$ were used, and $ L=1 $ was used in (iii) and (iv). The resulting super-windows and $\R$ are shown in figure \ref{fig:W}. 
	We initially suppose that the region $ \Omega $ is known. An estimation of this domain is presented later.
	
\subsection{ Bias quantification } 
	As discussed in section \ref{sec:Welch}, the bias in the estimation of $ \R $ and $ \C $ are related via Parseval's theorem, as in \eqref{eq:error}. 
	From \eqref{eq:bias}, a norm for the bias of the estimation obtained with a window pair of windows is constructed as 
	\begin{equation}\label{eq:anError}
		e^2 = \int \R^2(\tau)(1-W(\tau))^2 d\tau.
	\end{equation}

\begin{figure}[t!]
		\centering
		\includegraphics[scale=0.5]{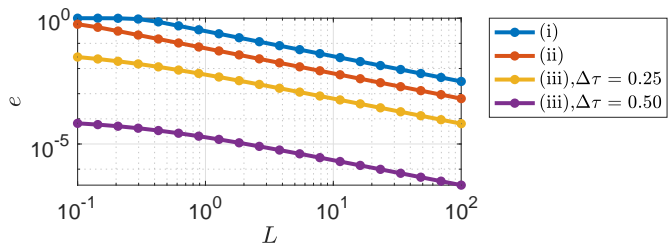}
		\caption{Bias norm \eqref{eq:anError}, obtained using different window pairs. }
		\label{fig:minimalerrors}
	\end{figure}

	Figure \ref{fig:minimalerrors} shows the bias norm using strategies (i)-(iii), where for (iii) values of $ \Delta \tau $ lower than the correct one were used to verify the robustness of the approach. Results using (iv) are analogous to (iii) and thus omitted. For all cases, the estimation mean goes asymptomatically to zero for large windows, indicating that all methods are unbiased in the limit of large window lengths. However, the error for a given, finite, window length varies drastically between the approaches.
	
	Compared to  (i), (ii) reduces the bias by a factor of $ 5 $. Using (iii), bias is a function of how well $ \Delta \tau $ is known. Even if $ \Delta \tau $ is underestimating by 50\%(20\%), the bias norm is $50(10^4)$  times smaller than the ones obtained with (i).  If the correct, or higher, value of $ \Delta \tau $ is used, the estimation becomes unbiased ($e=0$).

	\subsection{Numerical estimation of  $ R(\tau)$ and $\Omega$}
	 
	 In this section, $ \R $ and $ \C $ are estimated from a finite number of samples. The samples are obtained from different realizations and are thus uncorrelated. A unit window length is used for (i) and (ii), and $ L=1 $ for (iii) and (iv). 

\begin{figure}[t!]
	 	\centering
	 	\subfloat[Approach (i)]{
	 		\includegraphics[scale=0.5]{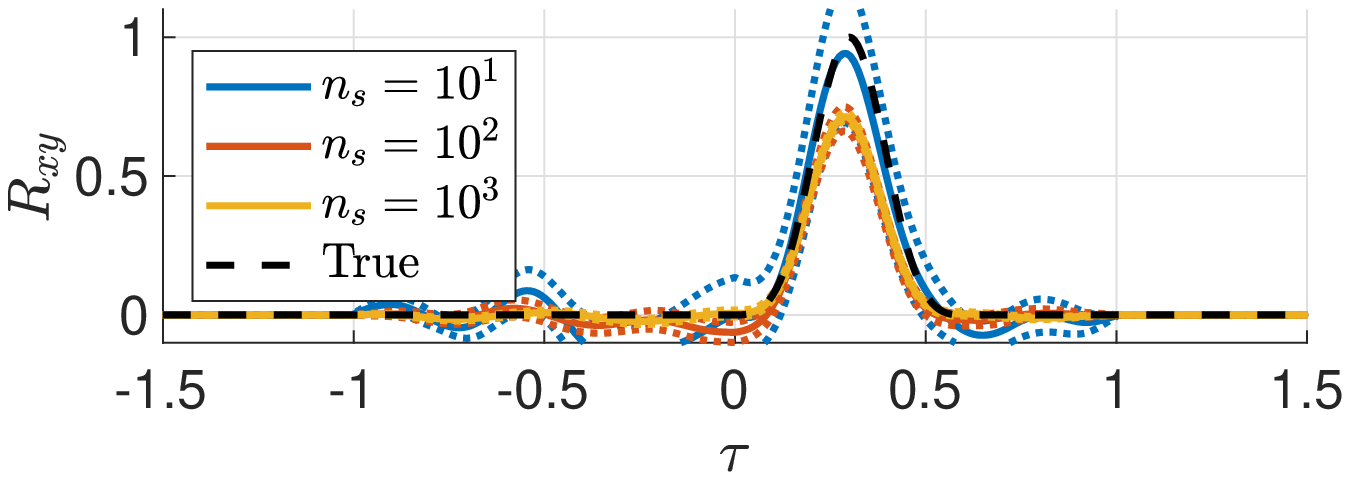}
	 	}
 	
	 	\subfloat[Approach (ii)]{
	 		\includegraphics[scale=0.5]{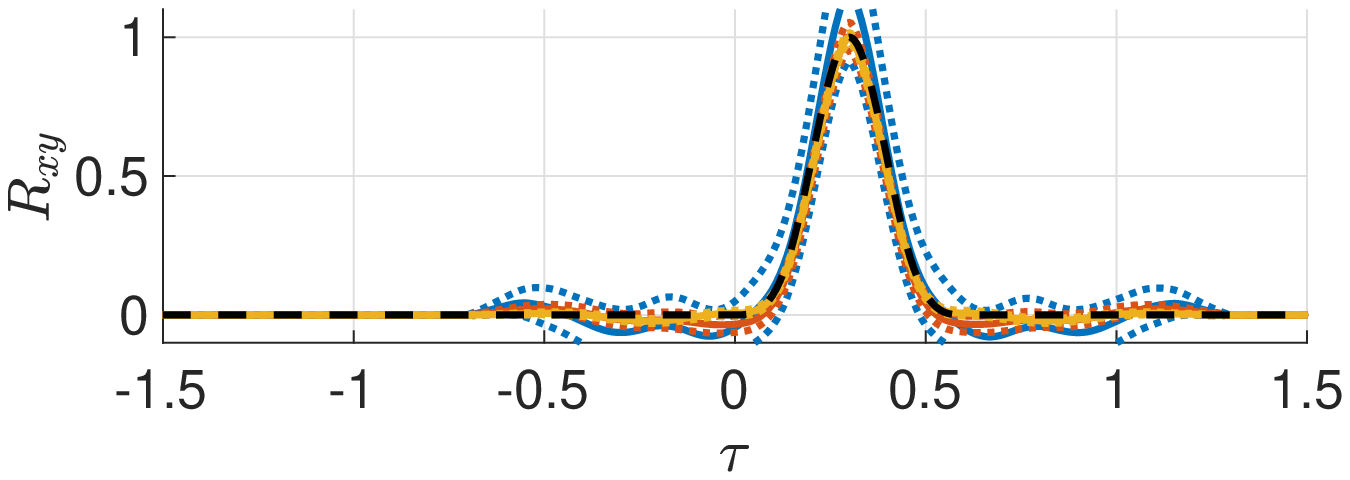}
	 	}
	 	\caption{Estimations of $\R$ using strategies (i) and (ii). Dashed lines correspond to one standard deviation of the estimate.}
	 	\label{fig:Rest}
	 \end{figure}

	We start by comparing estimates assuming that $\Omega$ is known. Figure \ref{fig:Rest} compares the estimates using (i) and (ii). Using (i), $ \R $ is consistently underestimated, while (ii) shows estimates that are visibly closer correct value. Results using (iii) and (iv) are visually similar to (ii).
	 	 
\begin{figure}
	 	\centering
	 	\includegraphics[scale=.5]{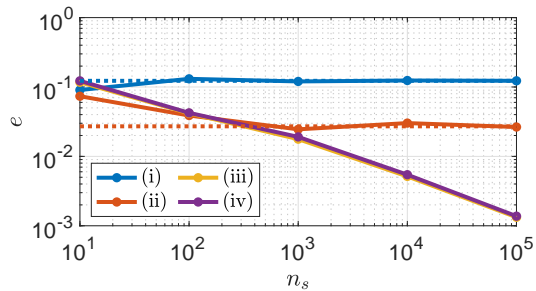}
	 	\caption{Estimation errors, defined as in \eqref{eq:error}, using different strategies. The dashed lines correspond to theoretical limits.}
	 	\label{fig:ensambleerrors}
	 \end{figure}

\begin{figure}[t!]
 	\centering 	
 	\subfloat[Approach (iii)]{
	\includegraphics[scale=0.5]{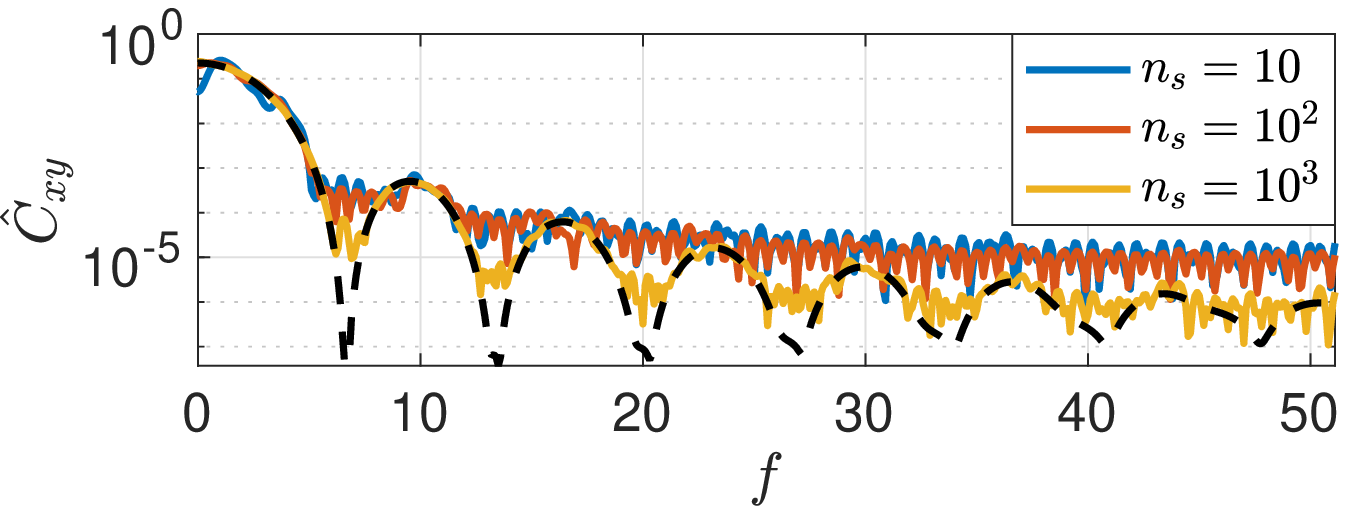}
	}

 	\subfloat[Approach (iv)]{
	\includegraphics[scale=0.5]{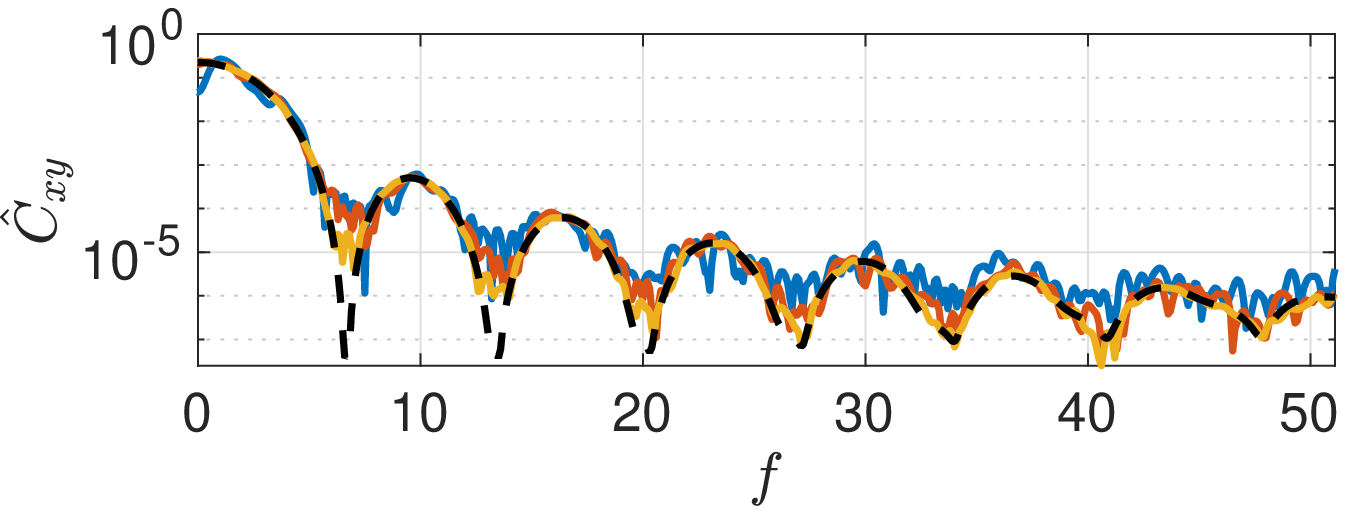}
	}
 	\caption{Estimations of $\C$, using a different number of samples. Dashed black lines indicate the true value of $ \C $. }
 	\label{fig:Cest}
 \end{figure}

	 Figure \ref{fig:ensambleerrors} quantifies the estimation errors using \eqref{eq:error}. Increasing the number of samples the estimation error approaches the bias given by \eqref{eq:anError}.
	 Error norms for  (iii) and (iv) are similar. However, figure \ref{fig:Cest} shows that (iv) better captures the behavior of $  \C $ at higher frequencies, due to the lower spectral leakage provided by the smooth super-window.  Both (iii) and (iv)  approaches show errors going to zero for a large number of samples, consistently with the fact that the two-windows approach is unbiased.
	 
\begin{figure}[t]
	  	\centering 
	  	\subfloat[]{\includegraphics[scale=0.5]{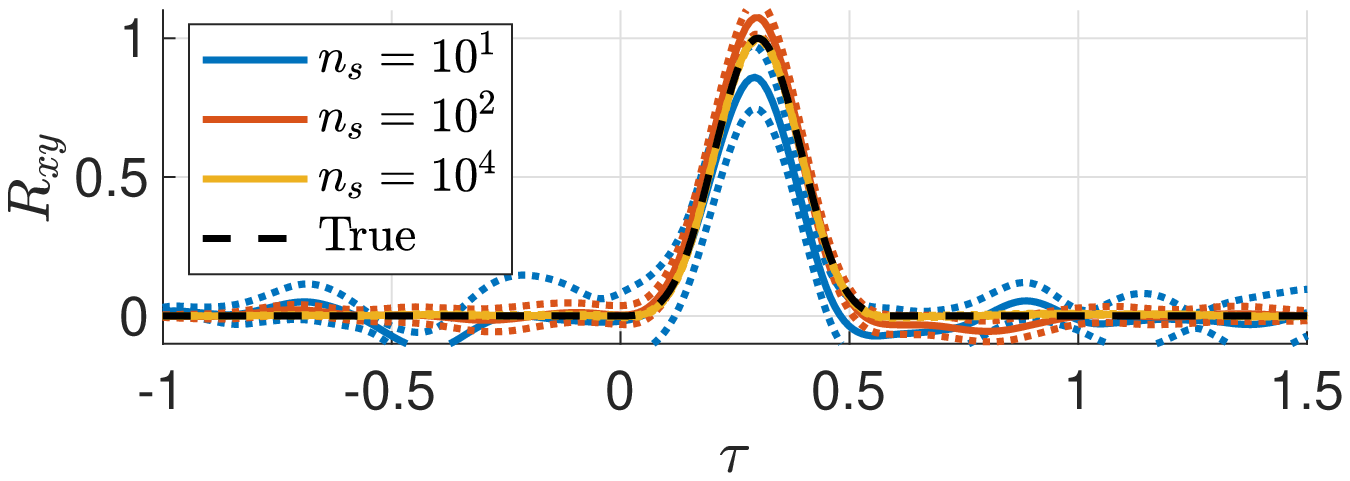}}
	  	
	  	\subfloat[]{\includegraphics[scale=0.5]{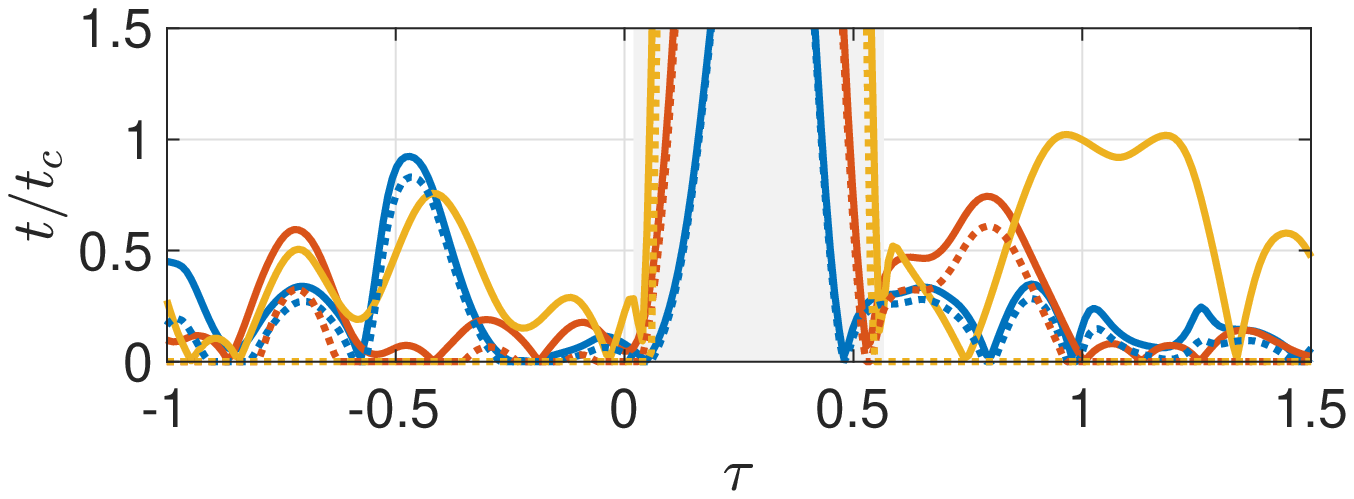}}
	  	\caption{In (a), estimations of $ \R $  (solid lines) and its one-standard-deviation bounds (dashed lines) are shown. The ratio between the $ t $-values using $ \epsilon=0 $ ($ 0.01 $) and the $ t_c $ are shown solid (dotted) lines in (b): $\tau$ is estimated to be in $\Omega$ if $t/t_c>1$. The region $ \Omega $ is highlighted in gray.  }
	  	\label{fig:omegaidenr}
    \end{figure}

    To estimate $\Omega$, the method proposed in section \ref{sec:estOmega} is used on uncorrelated samples. Figure \ref{fig:omegaidenr} compares the estimated and true values of $\R$, the estimation variance, and the corresponding t-values for three different numbers of samples.
    
    Although larger sample sizes improve the estimation $\Omega$, if no threshold is provided, stochastic noise can lead to false positives in the estimation of $\Omega$: see, for instance, the large value of $ t/t_c $ found for $ n_s=10^4 $ at $ \tau\approx 1 $. Specifying a tolerance value drastically reduces the probability of false positives while having little impact on the correct identification of $ \Omega $.
	  
	  
\section{Conclusion}\label{sec:conclusions}
    Welch's method was revisited, with an explicit representation of the expected value of the estimation obtained in the time-domain as the cross-correlation function modulated by a super-window. Analysis of the super-windows shows that the method is not biased for white-noise signals only.
	 
    A variation of Welch's method,  referred to as the two-windows approach, was proposed. The use of two different windows allows super-windows that are flat over the regions of interest to be obtained, thus providing unbiased estimations. The construction of the windows depends on knowledge of region $ \Omega $, which can be identified via simple and effective statistical tests.  Numerical experiments were presented, confirming the lack of bias in the proposed approach.
	 
    Besides providing an unbiased estimator, the approach proposed here also addresses the issue of window length. The slow convergence of the estimation bias with window length and the trade-off between window length and the number of samples makes unclear what is the optimal choice of window length.	 
	In the two-windows approach, the effective window length can be chosen considering computational costs only, being unbiased for any choice of length.
	 
\section*{Acknowledgment}
    The author would like to thank the discussions had with Peter Jordan, Andr\'e V. G. Cavalieri, Kenzo Sasaki, and Diego Chou.
	  
	\bibliographystyle{IEEEtran}        
	\bibliography{References}

	
	
\end{document}